\documentclass[aps, pra, 10pt, twocolumn, tightenlines, letterpaper, amsmath, amssymb, preprintnumbers, floatfix, longbibliography, nofootinbib]{revtex4-2}

\usepackage{dsfont}
\usepackage{amsmath}
\usepackage{amssymb}
\usepackage{physics}
\usepackage{tabu}
\usepackage{tabularx}
\usepackage{bm}
\usepackage{tikz}
\usetikzlibrary{shapes.geometric}
\usepackage{graphicx}
\usepackage{placeins}
\usepackage{textgreek}
\usepackage{multirow}
\usepackage{makecell}
\usepackage{soul}
\usepackage{diagbox}
\usepackage{xcolor}
\usepackage[normalem]{ulem}
\usepackage[export]{adjustbox}
\usepackage{float}
\usepackage{wasysym}

\newcommand{\bs}{\boldsymbol}

\usepackage[hypertexnames=false]{hyperref}
\hypersetup{
    colorlinks=true,       
    linkcolor=blue,          
    citecolor=blue,        
    filecolor=blue,      
    urlcolor=blue           
}

\newcolumntype{Y}{>{\centering\arraybackslash}X}

\usepackage{orcidlink}

\begin{document}

\begin{figure}
  \vskip -1.cm
  \leftline{\includegraphics[width=0.15\textwidth]{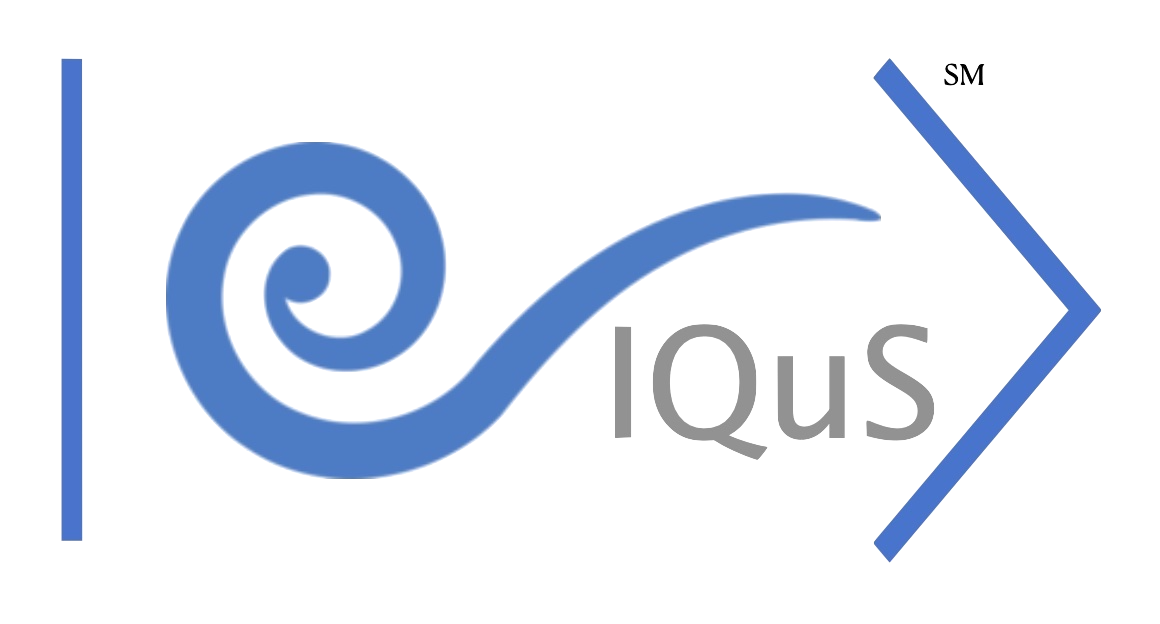}}
\end{figure}

\title{Dynamical Local Tadpole-Improvement in Quantum Simulations of Gauge Theories}

\author{Marc Illa\,\orcidlink{0000-0003-3570-2849}}
\email{marcilla@uw.edu}
\affiliation{InQubator for Quantum Simulation (IQuS), Department of Physics, University of Washington, Seattle, WA 98195, USA.}

\author{Martin J.~Savage\,\orcidlink{0000-0001-6502-7106}}
\email{mjs5@uw.edu}
\thanks{On leave from the Institute for Nuclear Theory.}
\affiliation{InQubator for Quantum Simulation (IQuS), Department of Physics, University of Washington, Seattle, WA 98195, USA.}

\author{Xiaojun Yao\,\orcidlink{0000-0002-8377-2203}}
\email{xjyao@uw.edu}
\affiliation{InQubator for Quantum Simulation (IQuS), Department of Physics, University of Washington, Seattle, WA 98195, USA.}

\preprint{IQuS@UW-21-100}
\date{\today}

\begin{abstract}
\noindent
We identify a new element in quantum simulations of lattice gauge theories, arising from spacetime-dependent quantum corrections in the relation between the link variables defined on the lattice and their continuum counterparts.
While in Euclidean spacetime simulations, based on Monte Carlo sampling, the corresponding tadpole improvement leads to a constant rescaled value per gauge configuration, in Minkowski spacetime simulations it requires a state- and time-dependent update of the coefficients of operators involving link variables in the Hamiltonian.
To demonstrate this effect, we present the results of numerical simulations of the time evolution of truncated SU(2) plaquette chains and honeycomb lattices in 2+1D, starting from excited states with regions of high energy density, and with and without entanglement.
\end{abstract}

\maketitle

\section{Introduction}
\label{sec:intro}
\noindent
Quantum simulations of lattice gauge theories (LGTs)
are central to improving predictive capabilities for 
the dynamics of key non-equilibrium processes in systems of fundamental particles~\cite{Banuls:2019bmf,Klco:2021lap,Bauer:2022hpo,Beck:2023xhh,DiMeglio:2023nsa,Bauer:2023qgm}.
These include 
the generation of the matter-antimatter asymmetry, 
the evolution of neutrinos and their role in the
production of heavy elements in supernova, 
and 
fragmentation and hadronization in high-energy collisions of matter.
While many of the techniques developed for classical simulations in Euclidean spacetime 
can be applied in
real-time quantum simulations, there are important modifications and new algorithms that remain to be identified, developed, implemented and verified.

A complete quantification of uncertainties in predictions from LGTs includes taking the continuum limit of ensembles of quantum simulations performed over a range of lattice spacings and volumes.
Lattice spacing artifacts result from the classical discretization(s) of a Hamiltonian or Lagrangian over the spacetime, and also from quantum fluctuations.
Recent works~\cite{Farrell:2024mgu,Ciavarella:2025zqf} have started investigating the 
impact of such effects on dynamical simulations of LGTs.
The classical lattice artifacts are well studied in Euclidean-space~\cite{Luscher:1984xn} and real-time~\cite{Moore:1996wn,Luo:1998dx,Carlsson:2001wp,Carlsson:2003rf,Carena:2022kpg,Gustafson:2023aai,Illa:2025dou} simulation frameworks, and can be mitigated order-by-order in powers of the lattice spacing using the Symanzik action(s)~\cite{Symanzik:1983dc}.
The plaquette operators defining the magnetic contribution to the Yang-Mills action, when expanded in terms of the finite lattice spacing, are renormalized by perturbative tadpole diagrams~\cite{Lepage:1992xa}.  
The ultraviolet (UV) divergence of these diagrams cancels their explicit lattice spacing coefficients to give a leading-order renormalization of the plaquette operators in 4D.  
One way to mitigate these unwanted quantum renormalizations is to implement tadpole improvement, which
involves numerically evaluating the expectation value of the plaquette operator in the gauge-field ensemble and  renormalizing the coefficient of 
the plaquette operators.\footnote{
In Euclidean spacetime, the expectation value of the plaquette operator is evaluated from the Monte Carlo sampled gauge-field configurations. This is efficiently performed using classical 
high-performance computers, being performed in parallel using lattice-shift operators on each gauge field in the ensemble. Another method
for estimating the tadpole coefficient is by computing the average of the link operator in Landau gauge~\cite{Lepage:1992xa}.}
In real-time Hamiltonian simulations, previous discussions of lattice-spacing artifacts~\cite{Luo:1998dx,Carlsson:2001wp,Carlsson:2003rf} 
considered
using the expectation value of the plaquette operator in the wavefunction but assumed it to be constant and uniform throughout a simulation. 
However, we point out that this expectation value depends on the strong-interaction environment surrounding the plaquette, and the mean-field value (hence improvement factor) is generally different for each plaquette, and is time dependent.
Not including the tadpole-improvement factors or using the expectation in the ground state of the Hamiltonian both lead to unquantified errors in subsequent predictions, 
including in the time-dependence and amplitude of observables.
In this work, we point out the need for time-dependent local tadpole improvement, and 
provide numerical evidence in the time evolution of 2+1D plaquette systems in pure SU(2) 
Yang-Mills lattice gauge theories.

\section{General Framework}
\label{sec:GF}
\noindent
The leading-order Kogut-Susskind Hamiltonian~\cite{Kogut:1974ag} describing non-Abelian LGTs, such as quantum chromodynamics (QCD), on a cubic lattice is a sum over the chromo-electric and chromo-magnetic contributions, 
\begin{align}
\hat H  = \ &  
\frac{g^2}{2 a^{d-2}} \
\sum_{ b, {\rm links}}
\ |  \hat {\boldsymbol E}^{(b)} |^2
\nonumber\\
& +
\frac{1}{2 a^{4-d} g^2} \sum_{i}
\left[
2 N_c - {1\over u_{0,i}^4} \left( \hat { \Box}_i^{\phantom{\dagger}} + \hat {\Box}_i^\dagger \right)
 \right] ,
\label{eq:QCDham}
\end{align}
where $g$ is the strong coupling constant, $a$ is the lattice spacing between adjacent sites, $d$ is the number of spatial dimensions, and $N_c$ is the number of colors.
The chromo-electric field  $\hat {\boldsymbol E}^{(b)}$ is summed over each link, 
with $b$ the SU($N_c$) adjoint index, 
and the chromo-magnetic field is approximated by the square plaquette operator $\hat { \Box}_i$, summed over the lattice
plaquettes, denoted by $i$ in Eq.~(\ref{eq:QCDham}).
This Hamiltonian can be improved by including terms parametrically suppressed by powers 
of the lattice spacing~\cite{Luscher:1984xn,Moore:1996wn,Luo:1998dx,Carlsson:2001wp,Carlsson:2003rf}
to permit the results of simulations to be closer to the continuum limit.
The quantity $u_{0,i}^4$ denotes the plaquette renormalization factor
(of the $i^{\rm th}$ plaquette)
that 
deviates from unity due to quantum fluctuations~\cite{Lepage:1992xa}.  
It arises from renormalizations of each link due to interactions with the fluctuating gluon fields.
In real-time evolution, these factors are generally time
dependent and location dependent.

Plaquettes are formed from products of link variables $U$ (Wilson lines) along a closed loop $\mathcal{C}$,
\begin{equation}
    \hat {\Box} = \prod_{k\in \mathcal{C}}U({\bs x}_k, {\bs e}_k) 
    \ ,
    \label{eq:plaq}
\end{equation}
and a general link variable from ${\bs x}$ to ${\bs x}+{\bs e}$ is defined as
\begin{equation}
U({\bs x}, {\bs e}) = \mathcal{P}\exp\left[ ig \int_{\bs x}^{{\bs x}+{\bs e}} dz_j A_j({\bs z})\right] 
\ ,
\end{equation}
where $\mathcal{P}$ denotes path ordering, 
$A_j$ is the vector potential and the summation is over spatial indices $j$. 
Perturbatively expanding this operator, in the coupling constant for a small lattice spacing, gives
\begin{align}
\label{eqn:Uexpand}
U({\bs 0}, {\bs e}_x) = &\ 
1 + i g a A_x({\bs 0}) 
+ i g {a^2\over 2} \partial_x A_x({\bs 0})
- {1\over 2} g^2 a^2 A_x^2({\bs 0})
\nonumber\\ 
& + \  \cdots \ ,
\end{align}
for a link in the $x$-direction.
The term quadratic in $A_x$
is naively suppressed by two powers of the lattice spacing,
but has a non-zero expectation value in a background gluon configuration.
In perturbation theory, 
the two-point function resulting from contracting the field operators in the 
$A_x^2({\bs 0})$ term,
evaluated in 4D Euclidean space, 
scales as 
\begin{equation}
\langle A_x^2({\bs 0}) \rangle \rightarrow 
\int {d^4q\over (2\pi)^4}\ {1\over q^2}
\ \sim\ {1\over a^2}
\ ,
\end{equation}
where we have used the lattice momentum cutoff 
(magnitude)
of $\pi/a$. 
In the continuum, it is UV divergent, canceling the explicit factor of $a^2$ in its coefficient 
in Eq.~\eqref{eqn:Uexpand},
and providing a finite renormalization of $U({\bs 0}, {\bs e}_x)$.
In lower dimensions, this provides a contribution that vanishes with the lattice spacing, 
but furnishes a finite  correction of the link variable away from unity at any finite $a$.

In the Hamiltonian framework, writing  
$A_x^2({\bs 0})$ in terms of annihilation and creation operators
leads to a background-dependent finite renormalization.
As an estimate of the deviation of the magnitude of the link from unity due to the strong-interaction background, 
the expectation value of the gauge-invariant plaquette operator (corresponding to four link operators in a (hyper-)cubic lattice)
is evaluated~\cite{Lepage:1992xa},
\begin{equation}
u_{0,i} = \left( 1 + \frac{1}{2 N_c} \langle\psi| \hat { \Box}_i^{\phantom{\dagger}} + \hat {\Box}_i^\dagger|\psi\rangle \right)^{1/4}
\ ,
\end{equation}
where $u_{0,i}$ is the link-renormalization factor appearing in Eq.~(\ref{eq:QCDham}).

\begin{figure}[t!]
    \centering
    \includegraphics[width=\columnwidth]{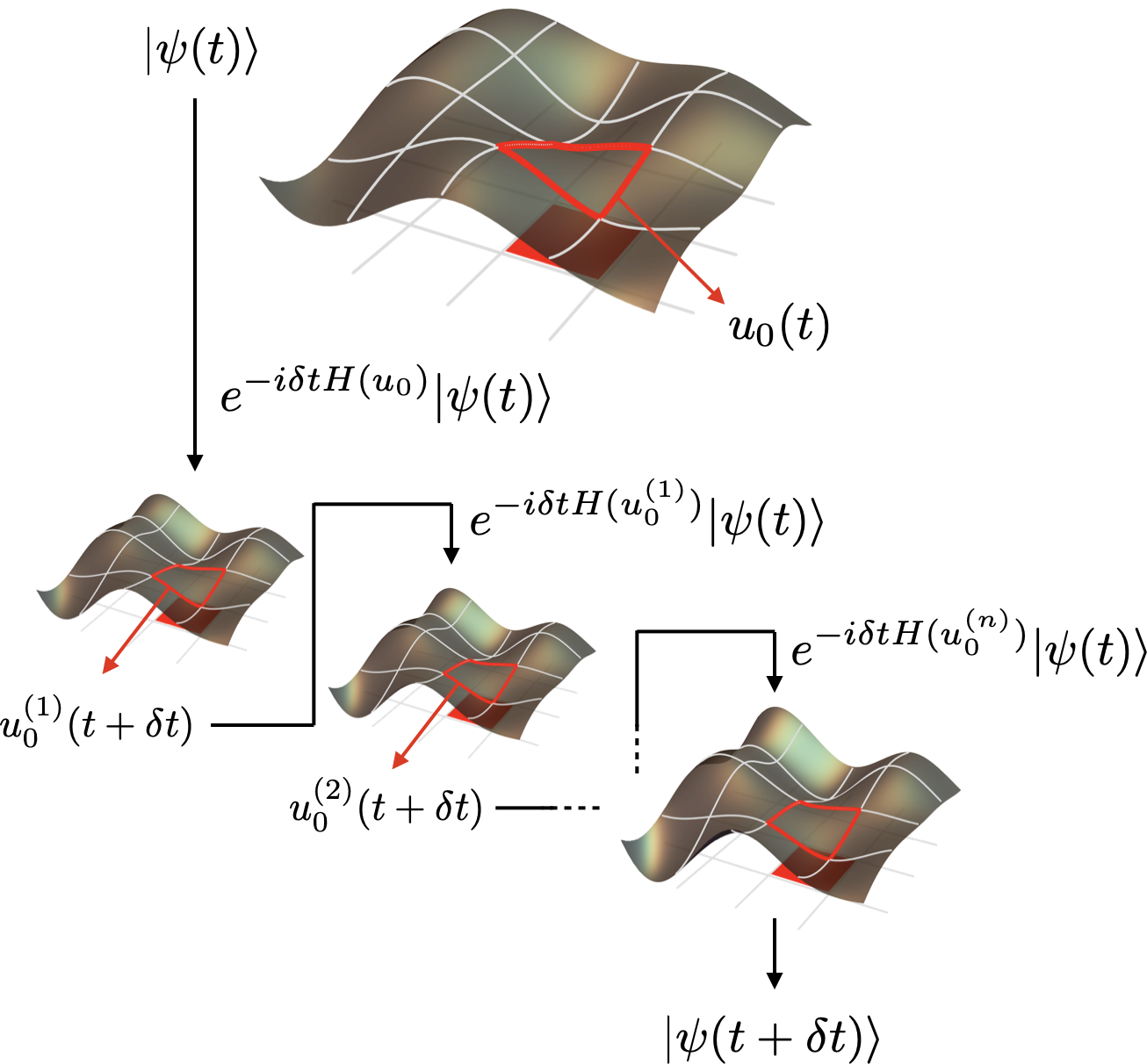} 
    \caption{Summary of the iterative dynamical local tadpole-improvement procedure, 
    where $\delta t$ is
    the size of the Trotter step used in the time evolution, and 
    $H(u_0) = H(\{ u_{0,i}(t) \})$ is a Hamiltonian of the form given in Eq.~(\ref{eq:QCDham}),
    a function of a set of spacetime-dependent mean-field values of the plaquette operator(s).
    }
    \label{fig:summaryALG}
\end{figure}
In Euclidean-space lattice QCD calculations, 
the $u_{0,i}$ are evaluated in the QCD vacuum ensemble,
which (on average) is 
(discrete)
translationally invariant, such that $u_{0,i}(t)=u_0$.
However, in real-time simulations from an arbitrary initial state, the $u_{0,i}(t)$ 
are time and position dependent.
At each time slice, the $u_{0,i}(t)$ need to be determined self-consistently, 
adding additional steps in the (Trotterized) time evolution.
For each Trotterized step forward in time, by an interval $\delta t$, 
the Hamiltonian is 
modified by the
expectation values of the plaquette operators 
in the new state.  This can be iterated to convergence, as displayed in Fig.~\ref{fig:summaryALG}.
A set of good starting conditions for the iteration of
$u_{0,i}(t+\delta t)$ are 
the converged values of $u_{0,i}(t)$.
For a sufficiently small $\delta t$, due to the observed exponential convergence of the iteration procedure, 
a precise value of $u_{0,i}(t+\delta t)$ can be obtained with a small number of iterations.
This introduces an element into the time evolution of quantum field theories
that had not been previously identified.

It is worth highlighting that, while the interacting vacuum of the theory remains 
(discrete)
translationally invariant after tadpole improvement, in general, the Hamiltonian is no longer translationally invariant during time evolution from an arbitrary initial state.  This is no surprise given that the state itself is generally not translationally invariant.  In light of this observation,  
implementations of quantum circuits that rely on underlying symmetries of the 
Hamiltonian need to be re-considered.

\section{Dynamical Local Tadpole Improvement}
\noindent
To illustrate the impact of spacetime-dependent tadpole improvement, 
we consider the time evolution of (a) 
chains of SU(2) plaquettes with periodic boundary conditions (PBCs)~\cite{Klco:2019evd,Hayata:2021kcp,ARahman:2021ktn,Yao:2023pht} and 
(b) honeycomb grids of SU(2) plaquettes with open boundary conditions (OBCs)~\cite{Muller:2023nnk,Turro:2024pxu}, both
truncated in the electric (irreducible representation) basis with $j_{\rm max}=1/2$.
This truncation in the field space significantly simplifies the action of the plaquette operators.\footnote{It is interesting to note that this construction is similar to the string of plaquettes in 
$Z_2$ LGT, which has recently been shown to exhibit scar states and other interesting thermalization properties~\cite{Hartse:2024qrv}.}

\subsection{SU(2) Plaquette Chains}
\label{sec:plaquettesu2}
\begin{figure}[t!]
    \raggedright
    \includegraphics[width=\columnwidth]{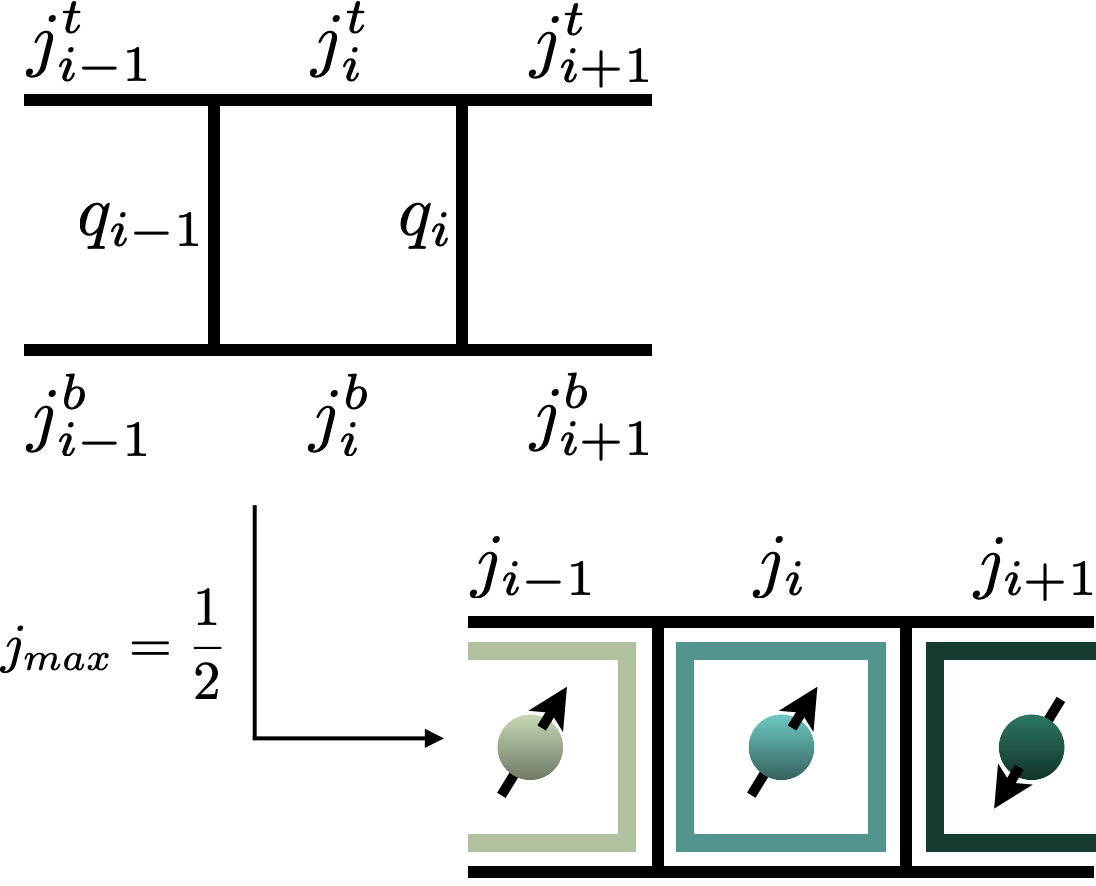} 
    \caption{The repeating unit in the mapping of the general SU(2) plaquette chain  to qubits.
    The SU(2) representation of the qubits assigned to the ``rails'' of the ladder are denoted by $j$, while those assigned to the ``rungs'' are denoted by $q$. After truncating to $j_{max}=1/2$, each plaquette is a two-level system, labeled only by its upper link $j_i$.}
    \label{fig:su2layout}
\end{figure}

\noindent
The repeating element in the 
mapping of the general SU(2) plaquette chain to qubits~\cite{Klco:2019evd} 
is shown in Fig.~\ref{fig:su2layout}.

The truncation in field space to $j_{\rm max}=1/2$ allows the general basis state of the plaquette chain to be written in terms of the representations of qubits associated with the upper ``rail'' of the chain
only,
$|j_0, j_1, j_2, ..., j_{L-1}\rangle$, for a plaquette chain of length $L$.\footnote{States constructed by applying various plaquette operators to the 
bare vacuum state have the same $j$ value in the upper and lower ``rails'' if $j_{\rm max}=1/2$. 
This is the well-known mapping of the corresponding $Z_2$ theory to a spin chain~\cite{Hayata:2021kcp,ARahman:2022tkr,Yao:2023pht}.}
Using this representation, the action of the plaquette operator on the $i^{\rm th}$ plaquette is implemented on this spin-chain with the controlled operator, 
\begin{equation}
\label{eqn:square_plaq}
\hat\Box_i =  
\hat\Lambda_0\hat X\hat\Lambda_0
 + 
{1\over 2}
\hat\Lambda_1\hat X\hat\Lambda_0
 + 
{1\over 2}
\hat\Lambda_0\hat X\hat\Lambda_1
 + 
{1\over 4}
\hat\Lambda_1\hat X\hat\Lambda_1
\ ,
\end{equation}
where we have suppressed the spatial indices, i.e., 
$\hat\Lambda_0\hat X\hat\Lambda_0 \rightarrow \hat\Lambda_{0,i-1} \otimes \hat X_i \otimes \hat\Lambda_{0,i+1}$.  
The coefficients of the four terms in this controlled-plaquette operator can be found
in Ref.~\cite{Klco:2019evd}.\footnote{The corresponding controlled-plaquette operators for SU(3) were introduced in Ref.~\cite{Ciavarella:2021nmj}.}${}^,$\footnote{The work presented in Refs.~\cite{Yao:2023pht,Ebner:2023ixq,Ebner:2024mee,Turro:2025sec} used the square-plaquette operator
\begin{equation}
\hat\Box_i =  
\hat\Lambda_0\hat X\hat\Lambda_0
 - 
{1\over 2}
\hat\Lambda_1\hat X\hat\Lambda_0
 - 
{1\over 2}
\hat\Lambda_0\hat X\hat\Lambda_1
 + 
{1\over 4}
\hat\Lambda_1\hat X\hat\Lambda_1
\ ,\nonumber
\end{equation}
which is equivalent to Eq.~\eqref{eqn:square_plaq} upon the basis change given by
\begin{equation}
U = \prod_{k} e^{-i\frac{\pi}{4}(-1)^k \hat{Z}_k\hat{Z}_{k+1}} \ .
\end{equation}
}
The projectors are defined as 
$\hat\Lambda_0=\tfrac{1}{2}(\hat I + \hat Z )$ and 
$\hat\Lambda_1=\tfrac{1}{2}(\hat I - \hat Z )$. 
From this, the chromo-magnetic contribution to the Hamiltonian given 
in Eq.~(\ref{eq:QCDham})  (with $N_c=2$ and $a=1$)
\begin{equation}
\hat H_B = 
{1\over 2 g^2}
\sum_{i=0}^{L-1}\ 
\left[ 
4 \hat I - {1\over u_{0,i}^4} \left( \hat \Box_i^{\phantom{\dagger}} + \hat\Box_i^\dagger \right)
\right]
\ ,
\label{eq:Magsu2Chain}
\end{equation}
can be written in terms of Pauli matrices.
Both the $j$ and $q$ links contribute to the chromo-electric Hamiltonian.
The later can be uniquely determined by projectors on adjacent $j$ qubits,
leading to 
\begin{equation}
\hat H_E = 
{3 g^2\over 8}
\sum_{i=0}^{L-1} 
\left[ 
2 \hat\Lambda_{1,i}
+ 
\hat\Lambda_{1,i}\hat\Lambda_{0,i+1}
+ 
\hat\Lambda_{0,i}\hat\Lambda_{1,i+1}
\right]
 ,
\end{equation}
while
the energy in a single plaquette is given by 
\begin{align}
\hat h_{E,i} = & \
{3 g^2\over 8}
\left[ 
2 \hat\Lambda_{1,i}
+ 
\hat\Lambda_{1,i}\hat\Lambda_{0,i+1}
+ 
\hat\Lambda_{0,i}\hat\Lambda_{1,i+1}
\right. \nonumber\\ 
& \left.
\qquad 
+ 
\hat\Lambda_{1,i-1}\hat\Lambda_{0,i}
+ 
\hat\Lambda_{0,i-1}\hat\Lambda_{1,i}
\right]
\ .
\end{align}

\begin{figure}[t!]
    \centering
    \includegraphics[width=\columnwidth]{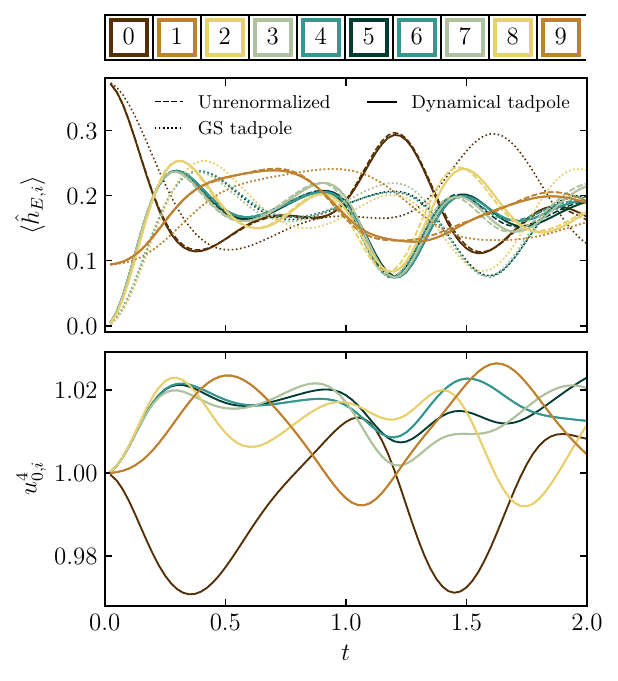}  
    \caption{The upper panel shows the 
    electric energy in each plaquette in the $L=10$ system 
    with $g=0.5$ as a function of time, starting from an initial tensor-product state of a single excited plaquette on the trivial vacuum, $|\psi_1\rangle$, as given in 
    Eq.~\eqref{eq:initialpsis_chain}.
    The solid line corresponds to the discretized time evolution with tadpole improvement iterated to convergence at each time step, $\delta t=0.025$.
    The dashed curves represent time evolution without tadpole improvement, and the dotted curves stand for using the interacting vacuum tadpole improvement. 
    The lower panel shows the corresponding tadpole improvement factors associated with the solid curves in the upper panel. 
    Plaquettes that share the same color evolve the same way due to the parity symmetry of the initial state.
    }
    \label{fig:su2PSI1}
\end{figure}

\begin{figure}[t!]
    \centering
    \includegraphics[width=\columnwidth]{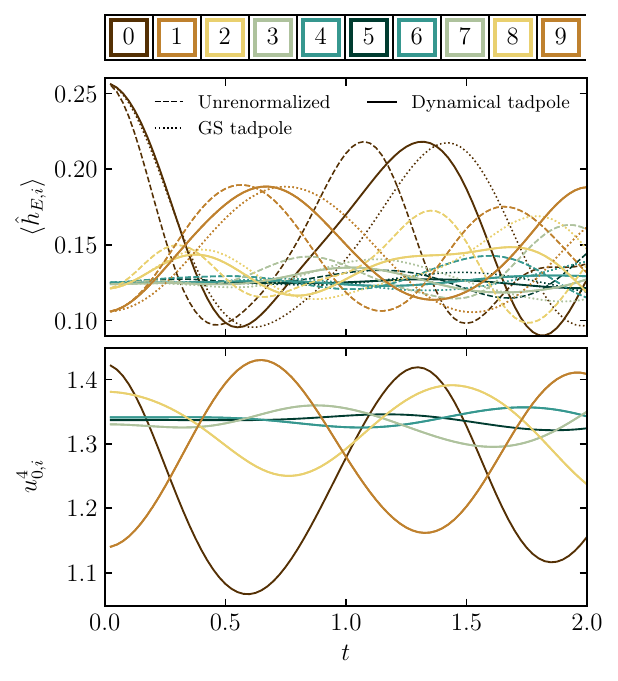} 
    \caption{Same as in Fig.~\ref{fig:su2PSI1}, starting from an initial entangled state  constructed by applying a single plaquette operator to the interacting vacuum, $|\psi_2\rangle$, as given in 
    Eq.~\eqref{eq:initialpsis_chain}.
    }
    \label{fig:su2PSI2}
\end{figure}

We present results obtained from a plaquette chain with $L=10$ and coupling of $g=0.5$.
Two different initial states are used to start the time evolution.
One is the trivial vacuum state, with a single plaquette excitation,  $|\psi_1\rangle$,
and the second is  the state obtained from a single application of the plaquette operator to the interacting ground state (vacuum) of the theory, $|\psi_2\rangle$,
\begin{align}
|\psi_1\rangle & = |1\rangle_0 \otimes |0\rangle^{\otimes (L-1)} \ ,
\nonumber\\
|\psi_2\rangle & = \mathcal{N}\ \hat\Box_0 |\psi_{\rm GS}\rangle_{u_0}
\ ,
\label{eq:initialpsis_chain}
\end{align}
where $\mathcal{N}$  is the appropriate normalization factor, and $|\psi_{\rm GS}\rangle_{u_0}$ 
is determined numerically with converged iteration in $u_0$.  
Translational invariance of the vacuum means that all of the $u_{0,i}$ in
$|\psi_{\rm GS}\rangle_{u_0}$
are identical.
For $g=0.5$, $u_{0}^4=1.33828$ and the energy density is $\epsilon_0=6.06346$ (in lattice units). For simplicity, 
the time evolution is not Trotterized, but a finite time-step is used to numerically define a unitary evolution operator from the full Hamiltonian.
The state $|\psi_1\rangle$ remains a tensor-product stabilizer state after exciting the single plaquette 
(in this spin basis). 
As such, it has vanishing entanglement for all multi-partite partitions.  
In contrast, as the ground state has non-trivial quantum complexity,\footnote{The ground state is entangled, and has non-zero magic density, 
e.g., 
with Stabilizer Renyi Entropies~\cite{Leone:2021rzd,Leone:2024lfr,Robin:2024bdz}: 
${\cal M}_1/L=0.445$ and
${\cal M}_2/L=0.337$ (both of which vanish in the trivial vacuum).
} 
the action of the plaquette operator to generate
$|\psi_2\rangle$ does not change this significantly, and this initial state remains
non-trivially entangled.
In both initial states, the action of the plaquette operator is such that the electric energy in 
3 of the plaquettes (i.e., $i=\{L-1,0,1\}$) differs from that in the rest of the $L-3$ plaquettes. 
This is because the action of the plaquette operator changes both $j$ and $q$ links.
Therefore, initially, there are 3 distinct values of plaquette electric energy, as seen in 
Figs.~\ref{fig:su2PSI1} and \ref{fig:su2PSI2}.

Starting time evolution from the  $|\psi_1\rangle$ initial state, both the electric energy density per plaquette and the tadpole-improvement factors are shown in Fig.~\ref{fig:su2PSI1}.
Over the time interval considered, 
the tadpole-improved evolution closely follows that of the unimproved Hamiltonian, 
consistent with $u_{0,i}\sim 1$.  However, if the tadpole improvement is taken from the ground-state wavefunction, significant deviations result (corresponding to the dotted curves in Fig.~\ref{fig:su2PSI1}).
This set of results is consistent with naive expectations.  Because of its off-diagonal structure, 
a non-zero matrix element of the plaquette operator requires a wavefunction with entanglement among relevant configurations.  Hence, $u_{0,i}=1$ for all plaquettes in the initial state, despite the distribution of 
electric energy density.

The results obtained for the system initially in the entangled state $|\psi_2\rangle$ are shown in Fig.~\ref{fig:su2PSI2}.
The tadpole-improved evolution is seen to differ significantly from that obtained using the unimproved Hamiltonian, and with the ground-state improvement factor.
This is reflected in the large relative changes to the dynamical local improvement factors (shown in the lower panel of Fig.~\ref{fig:su2PSI2}).
This initial improved Hamiltonian is not translationally invariant due to the structure of $|\psi_2\rangle$.
Because the eigenstates of the improved Hamiltonian have different energies from the unimproved or uniformly-improved Hamiltonians, observables such as the electric energy density become significantly different from each other on modest time scales.  
This provides an example demonstrating that while the classical effects of lattice spacing artifacts can be parametrically suppressed order-by-order, the quantum effects can become significant in time evolution over short time scales.

\subsection{SU(2) 2+1D Honeycomb Lattices}
\label{sec:honeycombsu2}
\noindent
We next examine the behavior of 2+1D SU(2) Yang-Mills gauge theory tessellated with honeycomb lattices and mapped to spin systems~\cite{Muller:2023nnk,Turro:2024pxu,Lee:2024jnt,Turro:2025sec,Illa:2025dou}.
This second implementation, while truncated in gauge space in the same way as the plaquette chains, probes the higher-dimensional aspect of dynamical local tadpole improvement in quantum simulations. 
The ${\cal O}(a^2)$-improved Hamiltonian of the honeycomb systems in terms of links and plaquettes is given by the sum of chromo-electric and chromo-magnetic contributions, where $a$ is the length of each edge of the honeycomb lattice plaquette and set to unity in the following.
The leading-order chromo-electric Hamiltonian is given by 
\begin{align}
H_E =
\frac{\sqrt{3} g^2 }{2} & 
\sum_{l,b} 
 \left( E^{(b)}_l \right)^2 
 \ ,
     \label{eq:HEHClinks}
\end{align}
where the summation over $l$ is for all links,
and the leading-order chromo-magnetic contributions is given by
\begin{equation}
    \hat{H}_B=\frac{2}{3\sqrt{3}g^2}\sum_{(i,j)}\left[4\hat{I}
    -\frac{1}{u^6_{0,(i,j)}}\left(\hat{\varhexagon}_{(i,j)}+\hat{\varhexagon}^\dag_{(i,j)}\right)\right] 
    \ ,
    \label{eq:HBHClinks}
\end{equation}
where the indices $(i,j)$ denote the spatial location of the plaquette.
The tadpole-improvement factor is given by 
\begin{equation}
u_{0,(i,j)} =  \left( 1 + \frac{1}{4} \langle\psi| \hat { \varhexagon}_{(i,j)}^{\phantom{\dagger}} + \hat {\varhexagon}_{(i,j)}^\dagger|\psi\rangle \right)^{1/6} 
\ .
\label{eq:u06HClinks}
\end{equation}

\begin{figure}[t!]
    \raggedright
    \includegraphics[width=\columnwidth]{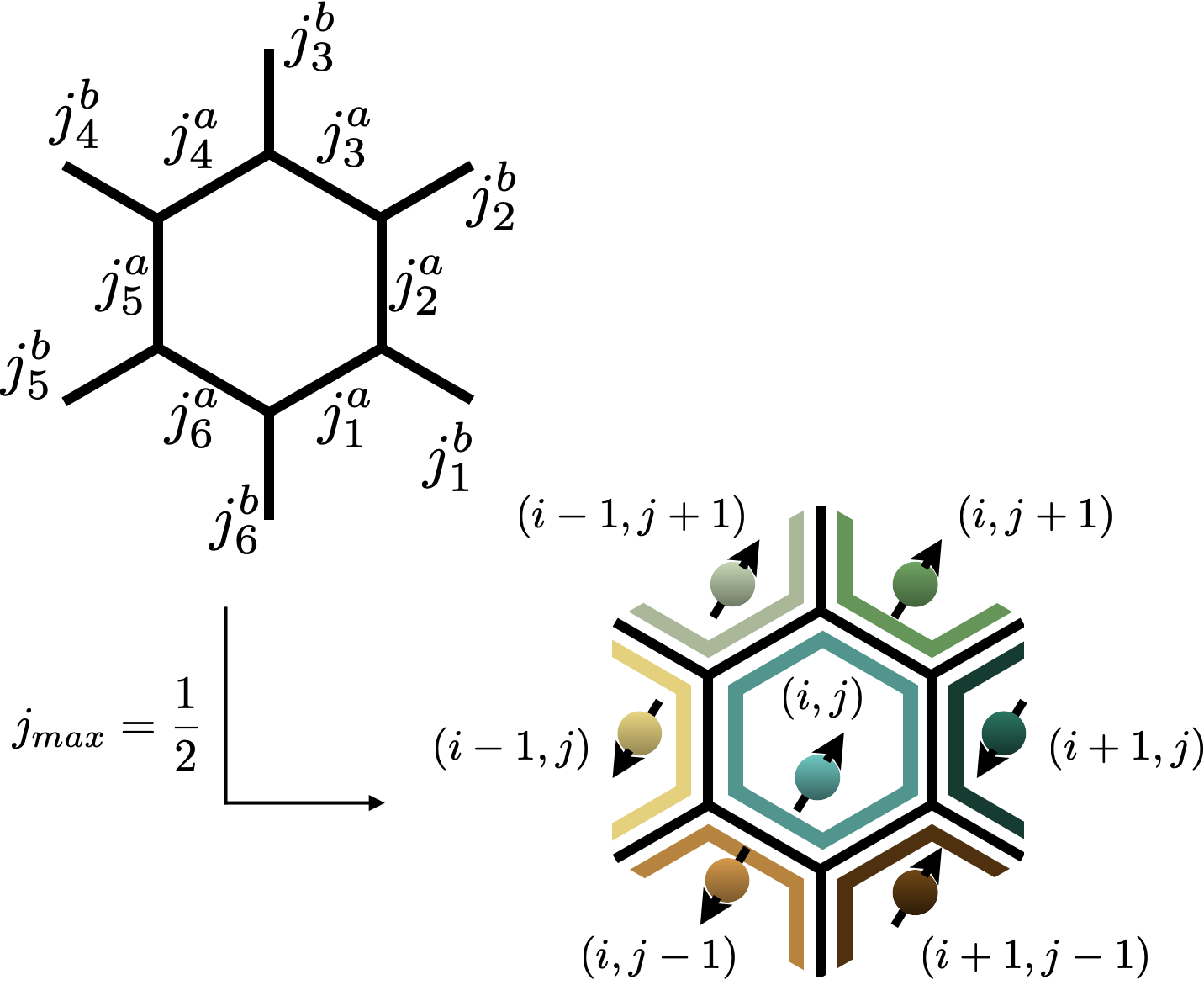} 
    \caption{
    The repeating unit in the mapping of the general SU(2) honeycomb lattice. The links belonging to the central cell are $j^a_i$, while the external legs controlling the action of the plaquette are $j^b_i$. As in the case of the linear chain of plaquettes, after truncating to $j_{max}=1/2$, each cell $(i,j)$ is mapped into a two-level system.}
    \label{fig:hexlayout}
\end{figure}

This system can be mapped to a system of spins with a gauge-space truncation 
of $j_{\rm max}=1/2$.
The plaquette operator in Eq.~\eqref{eq:HBHClinks} 
then becomes
\begin{equation}
    \label{eqn:plaq_all_+}
\hat{\varhexagon}_{(i,j)} = \hat{X}_{(i,j)}\prod_{K}\left[\left(\frac{1}{2}-\frac{1}{2\sqrt{2}}\right)\hat{Z}_{K}\hat{Z}_{K+1}+\frac{1}{2}+\frac{1}{2\sqrt{2}}\right] ,
\end{equation} 
with $K$ running over the nearest-neighbor spins around position $(i,j)$, 
\begin{align}
K =   \{ &
(i+1,j-1), (i+1,j), (i,j+1),  
\nonumber\\ 
& (i-1, j+1), (i-1, j), (i, j-1) \}
\ \ ,
\label{eq:hexNNlinks}
\end{align} 
see Fig.~\ref{fig:hexlayout}.\footnote{
The work presented in Refs.~\cite{Muller:2023nnk,Turro:2024pxu,Lee:2024jnt} define the hex-plaquette operator as
\begin{equation}
\label{eqn:plaq_some_-}
\hat{\varhexagon}_{(k,j)} = -\hat{X}_{(k,j)}\prod_{l}\left[\left(\frac{1}{2}-\frac{i}{2\sqrt{2}}\right)\hat{Z}_{l}\hat{Z}_{l+1}+\frac{i}{2}+\frac{1}{2\sqrt{2}}\right] \ .
\end{equation}
Equations~\eqref{eqn:plaq_all_+} and~\eqref{eqn:plaq_some_-} are equivalent, 
which can be seen by the basis change given by the unitary transformation from Ref.~\cite{Levin:2012yb},
\begin{equation}
U_{(i,j)} = \prod_K e^{-i\frac{\pi}{24}\left(3\hat{Z}_{(i,j)}\hat{Z}_K\hat{Z}_{K+1} - \hat{Z}_{(i,j)} - \hat{Z}_K - \hat{Z}_{K+1}\right)} \ .
\end{equation}
}

The chromo-electric field operators are matched to sums of spin-projector operators,
giving a total chromo-electric Hamiltonian of the form
\begin{align}
    \hat{H}_E & =
    \frac{3\sqrt{3}g^2}{4}
    \sum_{(i,j)}
    \hat{\Lambda}_{1,(i,j)}
    \nonumber\\
    &
    \times \left(3-\hat{\Lambda}_{1,(i+1,j-1)}-\hat{\Lambda}_{1,(i+1,j)}-\hat{\Lambda}_{1,(i,j+1)}\right)
    .
    \label{eq:hexHEspin}
\end{align}
When considering the electric energy density in a given plaquette, the sum of contributions from all six links are evaluated.   
This leads to 
\begin{equation}
    \hat{h}_E (i,j) = \frac{3\sqrt{3}g^2}{8}
    \left[
    \hat{\Lambda}_{1,(i,j)} \sum_K\hat{\Lambda}_{0,K}
    +
    \hat{\Lambda}_{0,(i,j)} \sum_K\hat{\Lambda}_{1,K}
    \right] ,
    \label{eq:HexEij}
\end{equation}
where the sum is over nearest-neighbor links, given in Eq.~\eqref{eq:hexNNlinks}.

\begin{figure}[t!]
    \centering
    \includegraphics[width=\columnwidth]{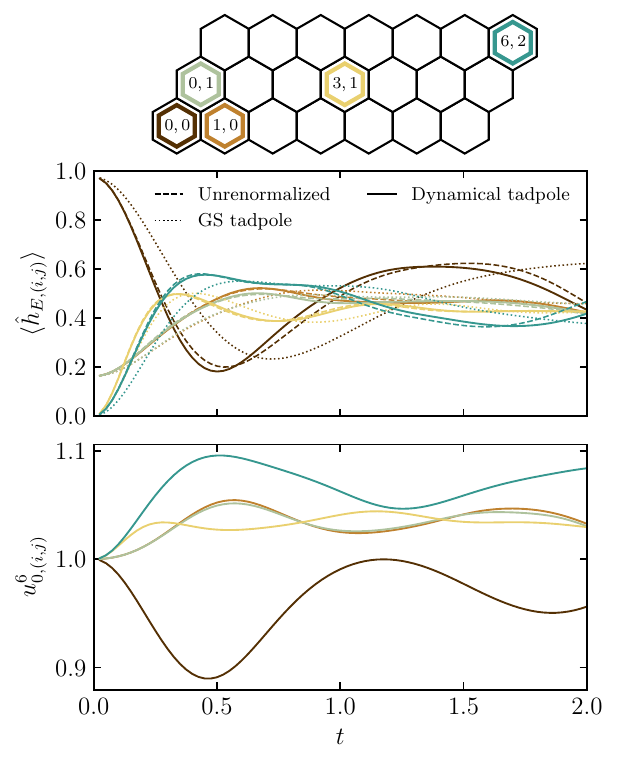} 
    \caption{The upper panel shows the 
    electric energy for a selection of the  hex-plaquette cells in a honeycomb lattice
    with $\{L_x,L_y\}=\{7,3\}$ and $g=0.5$ as a function of time, starting from an initial tensor-product state of a single excited plaquette on the trivial vacuum, $|\psi_1\rangle^{(2+1D)}$, as given in 
    Eq.~\eqref{eq:initialpsis}.
    The solid line corresponds to the 
    discretized time evolution with tadpole improvement iterated to convergence at each time step, $\delta t=0.025$.
    The dashed curves represent time evolution without tadpole improvement, and the dotted curves stand for  using the interacting vacuum tadpole improvement. 
    The lower panel shows the corresponding tadpole improvement factors associated with the solid curves in the upper panel. 
    }
    \label{fig:hexPSI1}
\end{figure}

\begin{figure}[t!]
    \centering
    \includegraphics[width=\columnwidth]{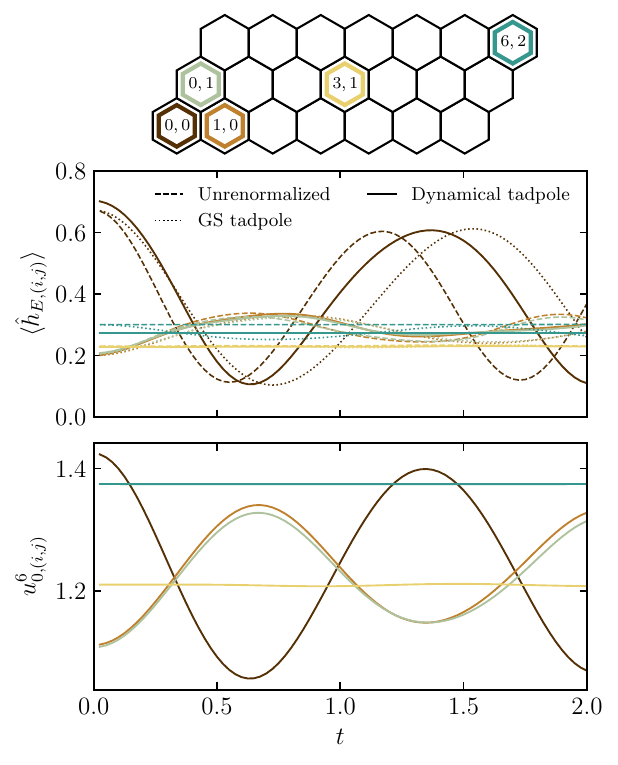} 
    \caption{Same as in Fig.~\ref{fig:hexPSI1}, starting from an initial entangled state  constructed by applying a single hex-plaquette operator to the interacting vacuum, $|\psi_2\rangle^{(2+1D)}$, as given in 
    Eq.~\eqref{eq:initialpsis}.
    }
    \label{fig:hexPSI2}
\end{figure}

To demonstrate the impact of dynamical tadpole renormalization in 2+1D simulations, we time evolve two different initial states, analogous to those used to initialize the plaquette chains, 
given in Eq.~\eqref{eq:initialpsis_chain},
\begin{align}
|\psi_1\rangle^{(2+1D)} & = |1\rangle_{(0,0)} \otimes |0\rangle^{\otimes (L_x L_y-1)} \ ,
\nonumber\\
|\psi_2\rangle^{(2+1D)} & = \mathcal{N}\ \hat{\varhexagon}_{(0,0)} |\psi_{\rm GS}\rangle_{u_0}
\ ,
\label{eq:initialpsis}
\end{align}
where $L_{x,y}$ denotes the number of hex-plaquettes in the $x$ and $y$ directions.
The boundary conditions we have used for the simulations are such that additional hexagons surround this active region, and the spins they are mapped to, are in the $|0\rangle$ state.\footnote{This is distinct from setting the electric field in the perimeter links into the $|0\rangle$ state.
}
Energy densities in the electric field associated with a selection of hex-plaquettes for a $7\times 3$ system are shown in Fig.~\ref{fig:hexPSI1} evolved from $|\psi_1\rangle^{(2+1D)}$, and in Fig.~\ref{fig:hexPSI2} for the evolution from $|\psi_2\rangle^{(2+1D)}$.
For this system, the values of the tadpole improvement factors for the ground state $|\psi_{\rm GS}\rangle_{u_0}$ are not uniform (due to using OBCs and $L_x\neq L_y$). For the highlighted cells in Figs.~\ref{fig:hexPSI1} and \ref{fig:hexPSI2}, the values are
\begin{align}
    u^6_{0,(0,0)} & = u^6_{0,(6,2)} = 1.37463 \ , \nonumber \\
    u^6_{0,(1,0)} & = 1.26440 \ , \nonumber \\
    u^6_{0,(0,1)} & = 1.25609 \ , \nonumber \\
    u^6_{0,(3,1)} & = 1.20984 \ ,
\end{align}
with an 
average
energy density of $\epsilon_0=4.97765$
(in lattice units).

The dynamical local tadpole improvements in the 2+1D honeycomb systems
exhibit the same features as in the linear plaquette chains. 
While the size of systems examined in this work are modest, 
they do not suggest that the effect(s) seen in 
the chains are diluted, or become less significant, in higher dimensions.
Tadpoles represent the nominally largest lattice artifact to the energies and time evolution,
and the higher-order lattice spacing artifacts, developed in Ref.~\cite{Illa:2025dou}, are expected to make parametrically smaller contributions.

\section{Summary and Outlook}
\label{sec:summary}
\noindent
Quantum simulations of emergent dynamic phenomena in strongly-interacting correlated systems are central to developing robust predictive capabilities for fundamental and applied physics.
In non-equilibrium dynamics, the relation between gauge-invariant operators and the definition of the Hamiltonian describing lattice gauge theories receives quantum corrections that vary across space and time, dependent upon the quantum state of the system at any given moment. 
These are self-consistently included into the dynamics by a local mean-field evaluation of the plaquette operator at each time step.
This is equivalent to re-scaling the link variables by an amount dependent on the local instantaneous gluon density.
We demonstrated this effect using  plaquette chains and honeycomb lattices of hex-plaquettes for 2+1D SU(2) gauge theory, both truncated in gauge space to $j_{\rm max} = 1/2$.
In particular, when the initial state is an entangled state, dynamical local tadpole improvement induces a bigger effect, compared to starting from a product (non-entangled) state.

While more sophisticated
studies remain to be performed, 
we expect this to be a particularly important effect in quantum simulations of highly-inelastic scattering 
processes, such as high-energy heavy-ion collisions with high-multiplicity final states.
Such initial states have high spatial inhomogeneity and high localized energy density 
(above the strong-interaction vacuum) within the two nuclei only.
The tadpole renormalization is expected to deviate significantly from that of the vacuum within the nuclei.
The high-multiplicity final state will, on average, have a much smaller energy density, with a distribution depending on fragmentation and hadronization in the event.
Finally, this effect is also expected to modify simulations of strong-field quantum electrodynamics, 
as the photon field itself will lead to modifications in the relation between plaquette operators and the magnetic field contribution to the Hamiltonian.

Recognizing the need for an iterative convergence to the correct
mean-field improved Hamiltonian that is space-time dependent is one thing, understanding how to modify present time-evolution quantum algorithms implemented on quantum computers is another.  
Naively, every relevant plaquette in the lattice is required to be measured in $|\psi\rangle$ at each Trotter step.
This, by itself, poses a challenge, since the plaquette is an off-diagonal operator (in the electric basis). For the examples shown above, simple protocols can be proposed, where the number of steps required to measure all plaquettes are independent of system size. However, for more generic cases (e.g., higher truncation of the gauge field or different basis), 
further investigations are required.
As the changes to the expectation values of the plaquette operators are expected to vary smoothly over a time scale consistent with the evolution of the wavefunction itself, if the expectation values are measured at each time step, and used in the following, this will provide an additive error to the overall leading-order Trotter error.
A more efficient implementation remains to be developed.
A path forward for implementing dynamic local tadpole improvement in LGT simulations includes increasing the truncation in the gauge field to ensure that the localization in gauge space is not enhancing the effect of the mean-field renormalization, 
to extend to simulations of SU(3) and higher gauge groups, 
to examine the behavior of larger systems in higher numbers of spatial dimensions, 
and to study the effect of shot- and device-noise in measuring these renormalization factors.

\begin{acknowledgments}
\noindent

We would like to thank Randy Lewis for helpful discussions regarding the classical implementation of tadpole improvement.
This work was supported, in part, 
by U.S.\ Department of Energy, Office of Science, Office of Nuclear Physics, InQubator for Quantum Simulation (IQuS)\footnote{\url{https://iqus.uw.edu/}} under Award Number DOE (NP) Award DE-SC0020970 via the program on Quantum Horizons: QIS Research and Innovation for Nuclear Science\footnote{\url{https://science.osti.gov/np/Research/Quantum-Information-Science}} 
(Martin, Xiaojun), and 
by the Quantum Science Center (QSC)\footnote{\url{https://www.qscience.org}} which is a National Quantum Information Science Research Center of the U.S.\ Department of Energy (Marc).
This work is also supported, in part, through the Department of Physics\footnote{\url{https://phys.washington.edu}} and the College of Arts and Sciences\footnote{\url{https://www.artsci.washington.edu}} at the University of Washington.
We have made extensive use of Wolfram {\tt Mathematica}~\cite{Mathematica}.
This research used resources of the National Energy Research Scientific Computing Center (NERSC), a Department of Energy Office of Science User Facility using NERSC award NP-ERCAP0032083.

\end{acknowledgments}

\bibliography{bib_main}

\end{document}